\documentclass[%
reprint,https://www.overleaf.com/project/6110737830fcd919fa3d7685
nofootinbib,
amsmath,amssymb,
aps,
]{revtex4-2}

\usepackage[format=plain,justification=centerlast]{caption}
\usepackage{subcaption}
\DeclareCaptionType{equ}[][]
\usepackage{floatrow}

\usepackage{graphicx}
\usepackage{arydshln}

\usepackage[table]{xcolor}
\usepackage{hhline}
\usepackage{hyperref}
\usepackage{amssymb}
\usepackage{amsthm}
\usepackage{amsfonts}
\usepackage{amsmath}
\usepackage{mathtools}
\usepackage{bm}

\usepackage{braket}
\usepackage[T1]{fontenc}
\usepackage[latin9]{inputenc}
\usepackage{color}
\usepackage{textcomp}
\usepackage{esint}
\usepackage{url}
\usepackage{natbib}
\usepackage[export]{adjustbox}
\usepackage{float}
\usepackage{dcolumn}

\usepackage[normalem]{ulem}

\date{\today}
\usepackage{lineno}

\begin{document}

\title{Direct measurement of the Husimi-Q function of the electric-field in the time-domain}

\author{Sho Onoe}
\email{sho.onoe@uqconnect.edu.au}
\affiliation{\emph{femtoQ} Lab, Department of Engineering Physics,
Polytechnique Montr{\'e}al, Montr{\'e}al, QC H3T 1J4, Canada}
\author{St{\'e}phane Virally}
\affiliation{\emph{femtoQ} Lab, Department of Engineering Physics,
Polytechnique Montr{\'e}al, Montr{\'e}al, QC H3T 1J4, Canada}
\author{Denis V. Seletskiy}
\affiliation{\emph{femtoQ} Lab, Department of Engineering Physics,
Polytechnique Montr{\'e}al, Montr{\'e}al, QC H3T 1J4, Canada}

\date{\today}

\begin{abstract}
{We develop the theoretical tools necessary to promote electro-optic sampling to a time-domain quantum tomography technique.
Our proposed framework implements detection of the time evolution of both the electric-field of a propagating electromagnetic wave and its Hilbert transform (quadrature).
Direct detection of either quadrature is not strictly possible in the time-domain, detection efficiency approaching zero when an exact mode-matching to either quadrature is reached.
As all real signals have a limited bandwidth, we can trace out the irrelevant sampling bandwidth to optimize the detection efficiency while preserving quantum information of the relevant signal. Through the developed understanding of the mode structure of the amplitude and Hilbert transform quadratures, we propose multiplexing and mode-matching operations on the gating function to extract full quantum information on both quantities, simultaneously.
The proposed methology  is poised to open a novel path toward quantum state tomography and quantum spectroscopy directly in the time domain.}
\end{abstract}

\maketitle

\section{Introduction}
Light can be used to detect and transport quantum information. In order to reveal this information, various quantum metrological techniques must be applied. All information can in fact be recovered via quantum-state tomography of the incoming light. The photocounting theory of Glauber, Kelley, and Kleiner \cite{PhotonCounting1963,PhotonCounting}, was one of the first attempts at a quantum mechanical interpretation of the detection of the radiation field.
However, even with unit quantum efficiency, this technique cannot extract the full quantum information of mixed or even pure states due to its insensitivity to phase/quadratures. The nominal work by Yuen and Shapiro \cite{YuenFirst} introduced the quantum mechanical interpretation of homodyne detection, which can detect both quadrature fields and their correlations at a desired frequency $\omega_0$. This interpretation has promoted our understanding of quantum field theory from a particle- to a field-based point-of-view incorporating the important role of the phase/quadratures. 

Although its statistics can be utilized to reproduce the Wigner function \cite{Collett1987, Vogel1989, WELSCH199963}, allowing full quantum-state reconstruction of the desired frequency, experimental implementations and extraction of quantum information have been met with difficulty due to the presence of noise. A leap in experimental progress was made with the introduction of the balanced-homodyne detection scheme \cite{Yuen:83}, capable of suppressing the technical noise in the self-homodyne method. Rigorous analysis of the (quantum) noise associated with the balanced homodyne detection \cite{Abbas1983,Schumaker1984,Yurke1985,Braunstein1990} together with the introduction of the spectral decomposition method \cite{YurkeWideband1985} paved the way toward quantum sensing in the frequency domain. These research led to the realization that efficient frequency domain homodyne detection of wavepackets require an understanding of the space-time character of the signal, to which the detection must be mode-matched. This insight led to the use of a quasi-monochromatic probe with a duration that is matched to the temporal duration of the signal, allowing efficient extraction of quantum information \cite{Aytur:92, Shapiro:97} from the state-under-study. Detection of non-classical states, starting with squeezed vacuum states \cite{Yurke1985,SqueezedLightLoudon,Slusher1985, Ling-An1986}, and the reproduction of their Wigner function \cite{Smithey1993} established balanced homodyne detection as one of the most reliable quantum sensing techniques in modern physics \cite{ScullyQuantumOptics, LoudonBook, MilburnQuantumOptics}.


Despite its success in the (quasi-)frequency domain, homodyne detection has not been successfully implemented for quasi-time-domain measurements \cite{Gulla2021}. For a monochromatic light, the quadratures are well-defined, and the transition between the two (e.g. $sin$ and $cos$) can be achieved via swap of parity or a simple time-delay, typically implemented in experiments via a change of group-velocity phase. In the time-domain, we can no longer swap between the two quadratures via a time-delay \cite{Stephane2019}: this simply changes the detection time and does not affect the parity of the measurement. As a result, the technical hurdle towards time-domain measurement is not only a broadening of the probe spectrum, but we must also correctly identify the quadrature fields in the quasi-time-domain, and seek the optimal spectral phases and amplitudes to mode-match the probe to those quadratures. 

Electro-optic sampling (EOS)\cite{CorrectionAlfred1999,Gallot1999,Huber2000,Kubler2004,Alexander2008}
has the ability, via nonlinearity, to isolate the electric-field from the Hilbert field (the quadrature orthogonal to the electric-field) at sub-cycle resolution. These features established EOS as one of the most reliable classical time-domain sensing techniques in the mid-infrared (MIR) frequency range, relevant for spectroscopic studies of molecular fingerprints \cite{Schweinberger:16}, semiconductors \cite{Kira2006} and various light-matter interactions \cite{Dorfman2016}. Its success in the classical regime has motivated its promotion to the quantum domain, where a team lead by Leitenstorfer achieved the first milestone towards quantum sensing: the direct detection of the MIR quantum vacuum in the time-domain with sub-cycle resolution \cite{Alfred_vacuum}. Since then, direct detection of terahertz quantum vacuum as well as its two-point correlations have been reported \cite{Buhmann_Faist} based on the EOS variants. However, there remain several milestones this technique must reach in order to establish itself as a reliable MIR quantum sensing technique in the time domain. Prominently among those are the identification of the quasi-time-domain quadratures and the ability to detect both field quadratures \cite{Stephane2019} and their correlations, and the experimental demonstration of extracting of quantum information from the field. 

Our Letter develops the theoretical framework for the former by accurately identifying the quadrature fields in the time domain and recasting the measurement protocols of EOS to be mode-matched to the detection of both simultaneously, hence achieving the direct measurement of the Husumi-Q function.  Gained insights of mode-matching are exemplified through a predicted improvement upon a recent experimental proposal for the detection of a quantum squeezed state\cite{Matthias}, which addresses the latter.

The remainder of this Letter is organized as follows: We first identify the quasi-time-domain quadratures as frequency-filtered time-domain electric-field and its Hilbert transform in Sec. \ref{SecEandH}. We then optimize the mode-matching and coupling efficiency in Sec. \ref{SecModeMatching}, to mitigate the introduction of undesirable vacuum noise and increase sensitivity to squeezing. In Sec. \ref{SecEOS} we promote EOS to efficiently couple to both the frequency-filtered electric field and its Hilbert transform simultaneously, giving the statistics of the Husimi-Q function which can utilized for full quantum-state reconstruction. We numerically demonstrate its effectiveness in quantum-state reconstruction of a squeezed state.

\section{Electric Field and its conjugate in the time-domain} \label{SecEandH}
Fields in (3+1) dimensions can be simplified to travelling-wave modes in (1+1) dimensions in the paraxial approximation, where they are considered to propagate along a single direction \cite{Blow1990,ScullyQuantumOptics,Sho2022UdW, Stephane2019}.
In such scenario we can decompose the electric field operator, $\hat{E}$, as 
\begin{equation}
\begin{gathered}
    \hat{E}_{\sigma}(t,x)= \int_{-\infty}^{\infty}\hspace{-3.5 mm} \mathrm{d} \omega \, E_{\omega}(t,x)( \, \hat{a}_{\omega,\sigma}) \, \text{, with}
\\
E_{\omega}(t,x) =-i \, \text{sign}(\omega)\sqrt{\frac{\hbar \lvert\omega\rvert c}{4 \pi n_{\omega}  A}}e^{-i \omega(t - \frac{n_\omega x}{c})} \, ,
\end{gathered}\label{EqnElectricField}
\end{equation}
where $A$ accounts for the transversal extension of the field, $n_\omega$ is the frequency-dependent index of refraction of the propagation medium, and the ladder operators $\hat{a}_{\omega,I}$ satisfy the canonical commutation relations $[\hat{a}_{\omega,\sigma},\hat{a}_{\omega',\sigma'}]=\text{sign}(\omega)\delta(\omega+\omega')\delta^\sigma_{\sigma'}$.
Here, we have adopted the notation where $\sigma$ refers to the polarisation degree of freedom, and $\hat{a}_{\omega,\sigma}=\hat{a}_{-\omega,\sigma}^{\dag}$.
The Hilbert quadrature ($H(\hat{E})(t)=\hat{H}(t)$) of the electric field is then decomposed as
\begin{equation}
\begin{gathered}
    {H}(\hat{E}_\sigma)(t,x)= \int_{-\infty}^{\infty}\hspace{-3.5 mm} \mathrm{d} \omega \, i \, \text{sign}(\omega)E_{\omega}(t,x) \, (\hat{a}_{\omega,\sigma}) \, .
\end{gathered}\label{EqnHilbertField}
\end{equation}

\begin{figure}[t]\includegraphics[width=\textwidth]{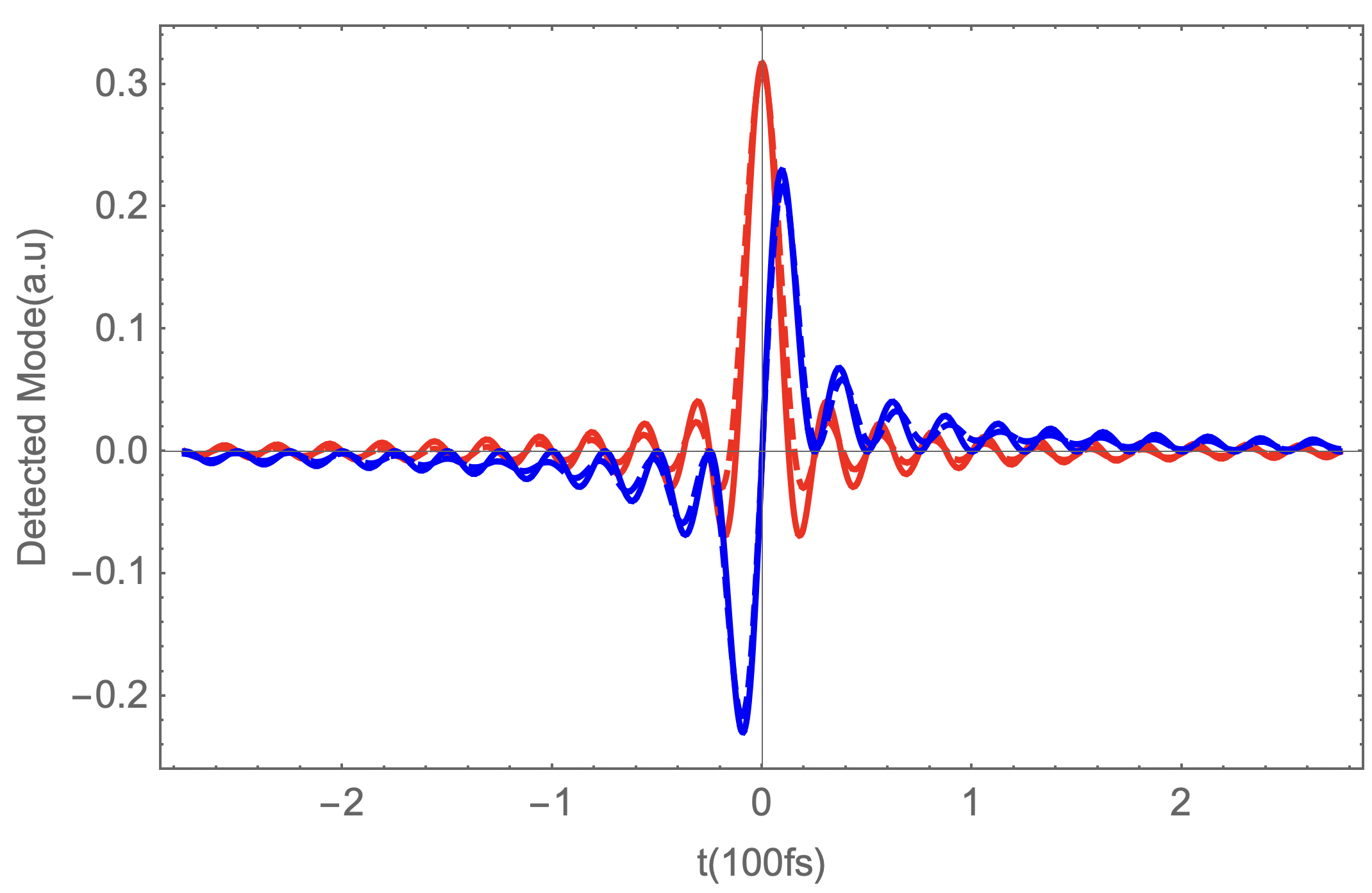}
\caption{The waveform of bandwidth-limited electric/Hilbert field are plotted in red/blue. These can be compared to the standard electric and Hilbert field which is a $\delta$-function and $1/t$ respectively. The red/blue dashed lines represent the coupled-mode via EOS.}
\label{FigEandHfield}
\end{figure}

The fundamental reason for the impossibility of measuring either quadratures strictly in the time-domain is the requirement for infinite detection bandwidth.
This mirrors the fundamental reason for the impossibility of measuring both quadratures strictly in the frequency-domain, namely the requirement for infinite detection time.
Any detection process will feature both finite bandwidth and a finite amount of interaction time with the field under probe. 

One may naively believe that approaching infinite bandwidth/time is desirable for the direct detection of quadratures in time/frequency-domain.
However, in a real set-up this leads to a coupling with fewer photons (e.g. there are fewer photons in a shorter time-window), reducing the coupling efficiency to the signal \cite{Matthias}: the preciseness in frequency/time comes at the cost of detection efficiency.
The detection efficiency approaches zero for the direct detection of quadrature in the time/frequency-domain. 

In the frequency-domain, the conventional approach is to impose a time-window based off information of the signal that is known \emph{a priori}.
As the signal does not exist outside this time-window, we do not lose any information by imposing this time-window to the probe.
In the same light, we can impose an appropriate bandwidth when extracting the information of the light signal.
The bandwidth-limited electric field and its Hilbert transform are defined as:
\begin{equation}
\begin{gathered}
    \hat{E}_{BL,I}(t,x)= \int_{-\omega_{m}}^{\omega_{m}}\hspace{-3.5 mm} \mathrm{d} \omega \, E_{\omega}(t,x) \hat{a}_{\omega,I} \, , \text{, with}
 \\
 \hat{H}_{BL,I}(t,x)= i \int_{-\omega_{m}}^{\omega_{m}}\hspace{-3.5 mm} \mathrm{d} \omega \,\text{sign}(\omega) E_{\omega}(t,x) \hat{a}_{\omega,I} \, .
\end{gathered}\label{EqnBandwidthLimitedEHfield}
\end{equation}
This simplification renders both $\hat{E}(t)$ and $\hat{H}(t)$ to be detectable (refer to Fig. \ref{FigEandHfield}), without losing any relevant information of the signal under study.
Furthermore, the finite bandwidth introduces the notion of mode-matching, which allows us to quantify how accurately we couple to these quadrature modes.

\section{Mode-matching and detection efficiency of EOS}\label{SecModeMatching}
Electro-optic sampling (EOS) is one of the most promising candidates in ultra-fast photonics for the detection of MIR electric field in the time-domain \cite{CorrectionAlfred1999,Huber2000,Kubler2004,Alexander2008}. It utilizes a $\chi^{(2)}$ based interaction between near-infrared (NIR) and mid-infrared (MIR) via sum- and difference- frequency generation.
By driving this interaction with a coherent probe-pulse with an envelope smaller than a single wavelength of the signal (i.e. sub-cycle with respect to the MIR signal), the sub-cycle quantum information is up-converted to the NIR.
\cite{Gallot1999}. This information is then analyzed via the ellipsometry scheme.
For a long time, experimental efforts have been made towards the direct detection of MIR electric field in the time-domain by utilizing an ultra-short pulse, reaching sub-cycle resolution under 6 fs \cite{Alfred_vacuum}.
This experimental effort for shorter pulses was motivated by the idea that the ideal probe is a Dirac-delta distribution.
The introduction of $\hat{E}_{BL}$ allows us to take a different narrative approach.

\begin{figure}
\centering
\includegraphics[width=\textwidth]{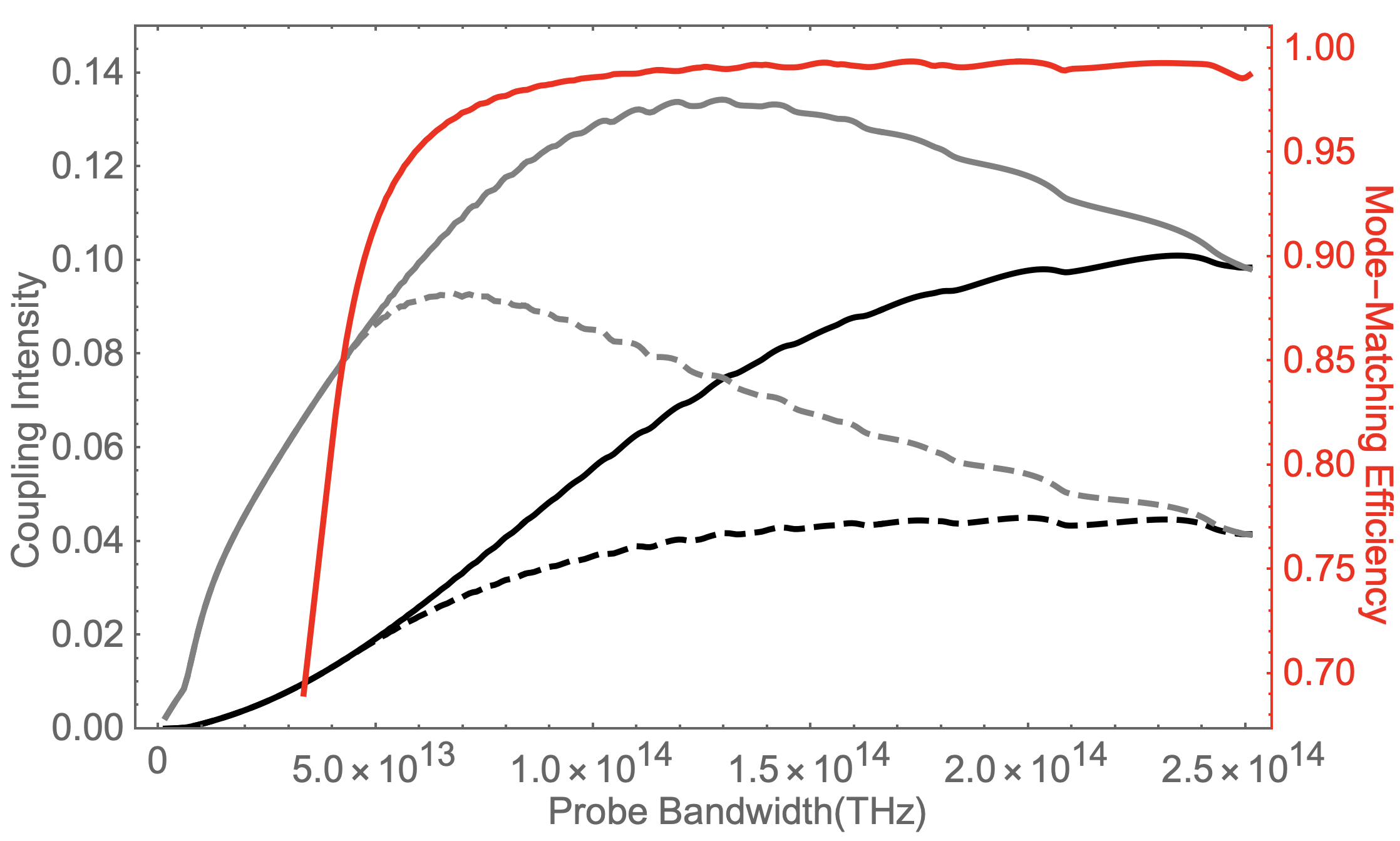}
\caption{We demonstrate the effect of bandwidth on coupling intensity and mode-matching efficiency. The black (dashed) line correspond to the constant photon number (bandwidth-limited) coupling intensity plots. The grey (dashed) line correspond to the constant intensity (bandwidth-limited), with coupling intensity plots. The red line correspond to the bandwidth-limited mode-matching efficiency as a function of bandwidth.}
\label{FigDetectionEfficiency}
\end{figure}

Let us consider a NIR signal with a central frequency of $\Omega_0=20$ THz.
The information of such pulse is generally limited to be under $2 \Omega_0$ \cite{IwoLocalization,Gulla2021}.
When this assumption is satisfied, we can neglect the quantum information of frequencies above $2 \Omega_0$ when analyzing this signal, and as a result we can utilize the bandwidth-limited electric field (refer to  Eq. \ref{EqnBandwidthLimitedEHfield}) with $\omega_{m}=2\Omega_0$.
Our goal is now tailored towards the detection of $\hat{E}_{BL}$. We can detect $\hat{E}_{BL}$ extremely accurately, with a mode-matching efficiency of $99\%$ by utilizing the EOS scheme with a sinc probe pulse. 
In Fig. \ref{FigDetectionEfficiency} we investigate the optimal bandwidth of the probe pulse to detect $\hat{E}_{BL}$ by analysing the effect of probe-pulse bandwidth on bandwidth-limited coupling intensity, $\theta_E$, and mode-matching efficiency, $\gamma_E$, which are defined as:
\begin{gather}
\gamma_{E}=:[\hat{E}_{BL},Tr_{BL}(\hat{\textbf{E}})]:/\sqrt{\theta_e \theta_E}\, ,
\\
\theta_E=:[Tr_{\omega_m}({\hat{\textbf{E}}}),Tr_{BL}({\hat{\textbf{E}})}]:\, ,
\\
\theta_e=:[\hat{E}_{BL},{\hat{E}_{BL}}]:\, .
\end{gather}
Here $\hat{\textbf{E}}$ is the detected electric-field via EOS (refer to Eq. (\ref{EqDetectedElectricField})), $Tr_{BL}[\hat{a}_{\omega}]=(1-\Pi((\omega)/(2\omega_m)))\hat{a}_{\omega}$, {$\Pi(x)=1 \, \forall-0.5\leq x\leq 0.5$, otherwise $\Pi(x)=0$,} and we have defined $:[,]:$ as the normal-ordered commutator, with $:[\hat{a}_{\pm|\omega|},\hat{a}_{\mp|\omega'|}]:=[\hat{a}_{|\omega|},\hat{a}_{|\omega'|}^{\dag}]$.  
 
 We find in Fig. \ref{FigDetectionEfficiency}, that when we keep the probe photon-number constant (at $5 \times 10^9$) the bandwidth-limited coupling intensity plateaus at about $4\%$ for large bandwidth. In comparison, the bandwidth-unlimited coupling intensity keeps increasing, implying that a larger bandwidth THz introduces additional noise from undesired frequencies. The amount of unwanted noise can be explicitly taken as the difference between to the bandwidth -limited and -unlimited detection efficiency.
 This implies we are using our resources in an inefficient manner, and we follow our analysis by considering the same plot, but with a constant intensity of the probe pulse motivated by the damage threshold of EOX crystal.
 In this scenario, a smaller bandwidth (roughly $70$ THz) probe pulse shows a significant increase in coupling intensity, reaching roughly $9\%$ in coupling intensity.
It is noted that increasing the photon number does not change the ratio between the wanted and unwanted noise.
 
 We also plot the bandwidth-limited mode-matching efficiency, which shows that a mode-matching of $99\%$ is achieved when the full-bandwidth is utilized.
 We constrain ourselves to a mode-matching efficiency that is at least of that achieved by the full-bandwidth.
 Under this condition, the optimal coupling intensity of $8\%$ is achieved at around $120$ THz.
{By considering the appropriate spectral support we suppress the admixture of unwanted vacuum fluctuation and by efficiently mode-matching our probe specifically to the signal in question, we} improved the detection efficiency \cite{Matthias} by roughly two folds without the cost of accuracy.
\section{Full Tomography}\label{SecEOS}

\begin{figure*}[t]
\centering
\begin{subfigure}{.5\textwidth}
  \centering
  \includegraphics[width=1\linewidth]{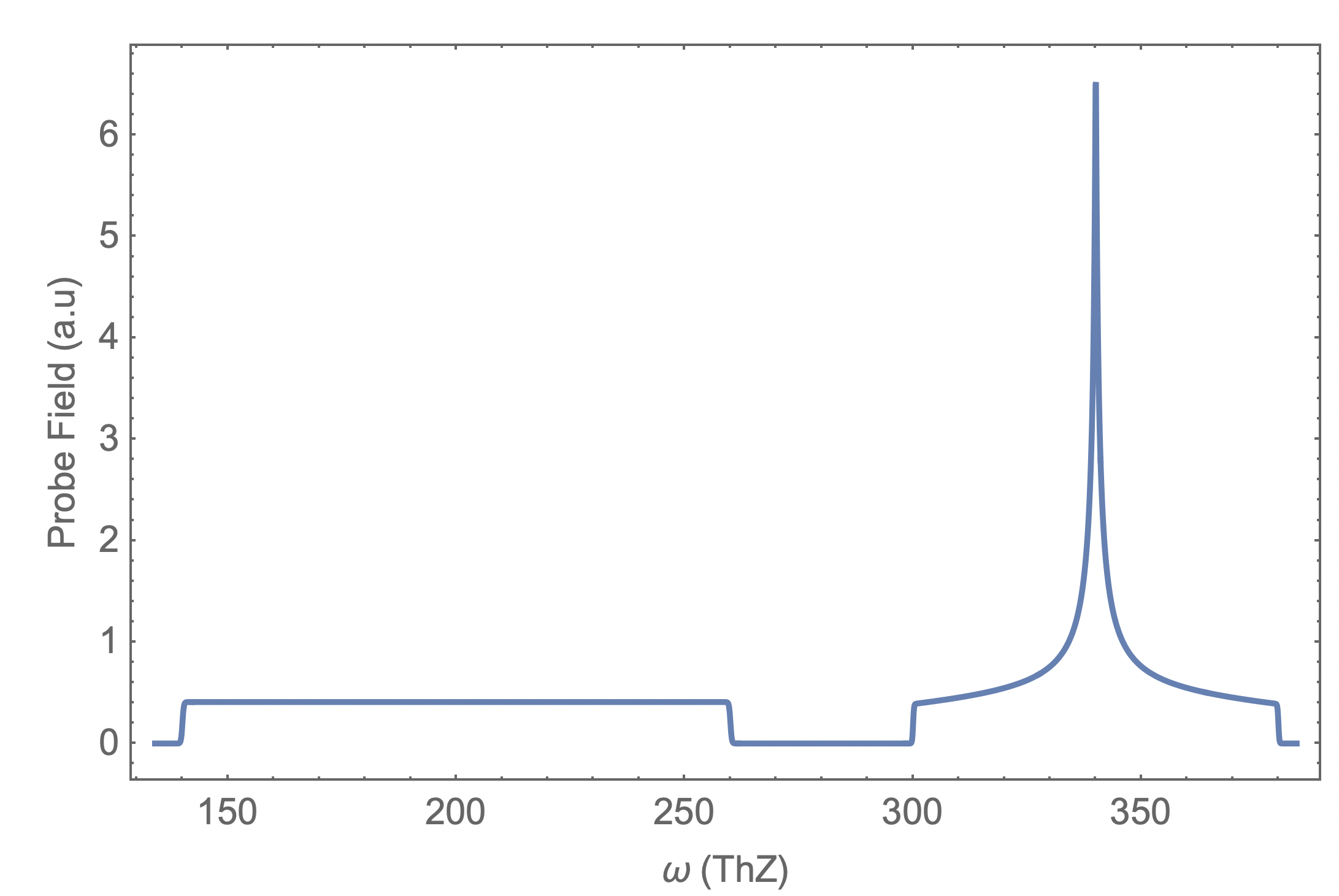}
  \caption{Spectral decomposition of probe-pulse electric-field }
  \label{FigProbePulse}
\end{subfigure}%
\begin{subfigure}{.5\textwidth}
  \centering
  \includegraphics[width=1\linewidth]{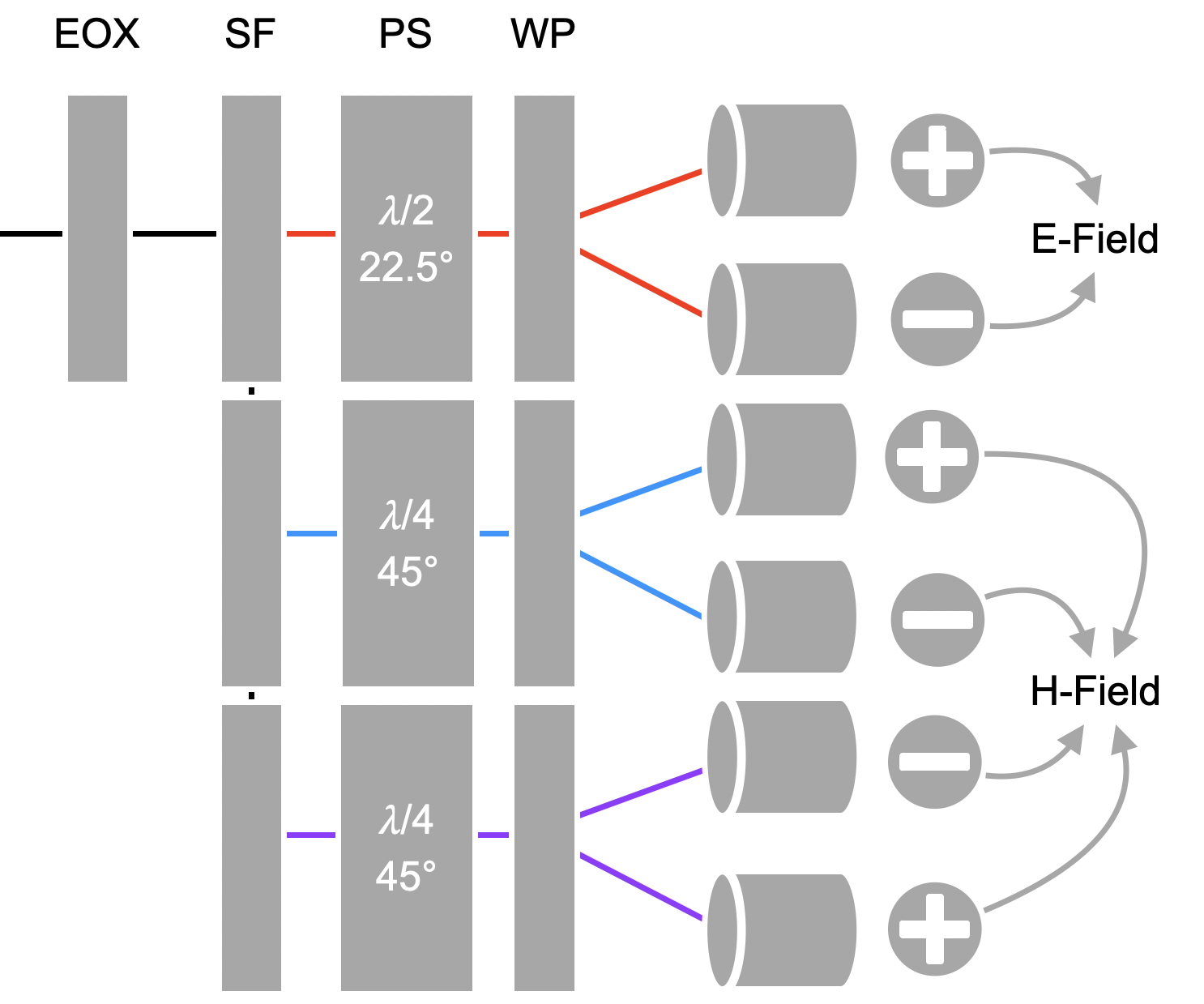}
  \caption{Experimental set-up}
  \label{FigExperimental}
\end{subfigure}
\caption{a) We show the probe pulse electric-field that is utilised to drive the EOX interaction. The signal {below/above $275$ THz is dedicated towards the detection of the electric-field/Hilbert-field quadrature.} b) We present a novel schematic set-up for the simultaneous detection of the electric- and Hilbert- quadrature via EOS. The probe pulse goes into the EOX crystal 
in the Z-polarisation 
, while the signal in the S-polarisation 
. This is followed by a spectral filter beam-splitter
, splitting the three frequency regimes (red for below $275$ THz, purple for above $340$ THz and blue for frequencies in-between). The first port has a half-waveplate in the 22.5 degrees from the z-polarisation. The second and third port has a quarter-waveplate 45 degrees from the z-polarisation. This is followed by a Wollaston prism which separates the S- and Z- polarisation fields, where the intensity difference between the polarisations are recorded.}
\label{FigExpBoth}
\end{figure*}

One of EOS's strength over homodyne detection is its ability to isolate the electric-field from the Hilbert quadrature regardless of the probe pulse's phase. This introduces the notion of an absolute phase \cite{Stephane2019,Sho2019universal} of the signal and we lose the meaning of relative phase. 

{In contrast, for homodyne detection we are agnostic to whether we are coupling directly to the electric-/Hilbert- field, but we obtain information on the relative phase between the signal and probe.} In homodyne detection, the relative phase is altered by a time-delay to the signal. {For EOS, this method cannot be used to alter the absolute phase, as it simply shifts the time that is sampled without affecting the quadrature (e.g. $\hat{E}_{BL}(t)\rightarrow \hat{E}_{BL}(t')$).} In the following sub-section we identify how EOS can be altered to detect the Hilbert transform.

\subsection{Detection of the Hilbert transform} 
\label{SecDetectionH}
{When we are analysing the electric field, we can obtain its localized property in the time-domain by a mode with an even parity. Its Hilbert transform is represented by a mode with odd parity which is a delocalized property of the electric-field (refer to Fig. \ref{FigEandHfield}).}
From the motivational stand point of view, EOS aimed at extracting the local property of the electric-field and therefore fails to extract the Hilbert transform.
There are two fundamental pillars of EOS that needs to be removed for the detection of the Hilbert transform: its ability to detect phase-locked (in frequency) electric-field mode and localized modes.
The prior can be resolved by the introduction of spectral filters, and adding a different phase contribution to different frequencies.
Changing from a flat phase with respect to the electric field to an odd-phase contribution (positive above the central frequency, negative below the central frequency) is desirable for the detection of a odd-function (e.g. the Hilbert transform).
As the Hilbert quadrature is a delocalized quantity of the electric-field, we benefit from utilizing a delocalized probe-pulse with a strong central frequency contribution (refer to Fig. \ref{FigProbePulse} for the exact waveform).
In the App. \ref{AppHilbertTransform}, we analyse the effect of bandwidth on detection efficiency and coupling strength, and find that a peak mode-matching of $99.9\%$ is achieved at $80$ THz bandwidth with a coupling intensity of roughly $2\%$.

\subsection{Simultaneous detection of both field quadratures}
{So far, we have established a mean to detect the field and its Hilbert transform independently}.
We can conduct both simultaneously via multiplexing and assigning different frequency regime to the detection of either.
We demonstrate this experimental setup in Fig. \ref{FigExperimental}.
In this paper we evaluate the effect of the optical elements via the Heisenberg picture.
Driving the EOX crystal utilizing a strong coherent signal $\alpha_z(t,x)$ leads to a Hamiltonian of the following form \cite{Sho2022UdW}:
\begin{equation}
\hat{H}_{\chi}(t)=\int_{-\infty}^{\infty} \mathrm{d}x \lambda {\alpha}_z(t,x) \hat{E}_S(t,x)\hat{E}_S(t,x)\Pi\left(\frac{x}{L}\right),    
\end{equation}
where $\lambda=\frac{A\epsilon_0 d}{2}$, with the coupling constant $d=-n^{4}r_{41}$, expressed in terms of the electro-optic susceptibility coefficient, $r_{41}$, and the refractive index of the crystal, $n$ \cite{Andrey, Sho2022UdW}.
We consider, without the loss of generality, a zincblende-type crystal with a length of $L=7 \, \mathrm{\mu m}$.
Its electro-optic coefficient is taken as $r_{41}=4~\mathrm{pm/V}$ (for the particular case of ZnTe).
The refractive index, $n_{\Omega}$, varies only slightly (from 2.55 to 2.59) in the MIR \cite{Andrey}.
We utilize a fit for the refractive index in the NIR frequency range $n_{{\omega}}$ \cite{Marple1964,Sho2022UdW}.
We compute the effect of this Hamiltonian by utilizing a unitary correction \cite{Sho2022UdW} to second order perturbation \cite{TLMGuedesPerturbation, Boyd}.
This is followed by a spectral filter, splitting the output into three frequency components:
\begin{equation}
    \begin{gathered}
    \hat{N}_{i,\sigma}=\int^{\omega_{i,max}}_{\omega_{i,min}} \mathrm{d}\omega \; \hat{a}_{\omega,\sigma}^{\dag}\hat{a}_{\omega,\sigma}\, ,
\end{gathered}\label{EqPhotonNumber}
\end{equation}
Each port goes through either a half- or quarter- waveplate:
\begin{equation}
    \hat{U}(\theta_i,\phi_i)=\hat{e}^{i \phi_i (\cos(\theta_i)\hat{N}_{i,S}+\sin(\theta_i)(\hat{N}_{i,Z})}\, .
\end{equation} 
Details of $\theta_i$ and $\phi_i$ can be found in Fig. \ref{FigExperimental}.
In each branch, the waveplate is followed by a Wollaston prism and a balanced pair of photo-detectors. The E-field is given by
\begin{equation}
\hat{\textbf{E}}(t)=\frac{\hat{N}_{z,1}'-\hat{N}_{s,1}'}{\sqrt{\braket{\hat{N}_{z,1}'+\hat{N}_{s,1}'}
}}\, ,\label{EqDetectedElectricField}
\end{equation}
and its Hilbert transform,
\begin{equation}
\hat{\textbf{H}}(t)=\frac{\hat{N}_{z,2}'-\hat{N}_{s,2}'}{\sqrt{\braket{\hat{N}_{z,2}'+\hat{N}_{s,2}'}}}+\frac{\hat{N}_{s,3}'-\hat{N}_{z,3}'}{\sqrt{\braket{\hat{N}_{z,3}'+\hat{N}_{s,3}'}}}\, .
\end{equation}
Simultaneous detection of both allows the detection of the Husumi-Q function. In App. \ref{AppDifferentMethods}, we discuss two other promising schemes that can be implemented for signals with larger bandwidth.
One of these scheme is analogous to the method discussed in this section, utilizing a beam-splitter instead of a spectral filter (refer to App. \ref{AppDifferentMethods1}) , while the other can be utilized for full tomography with an arbitrary phase (refer to App. \ref{AppDifferentMethods2}), which can be utilized to extract the Wigner function.

\section{Numerical Results}
\begin{figure}[t]\includegraphics[width=\textwidth]{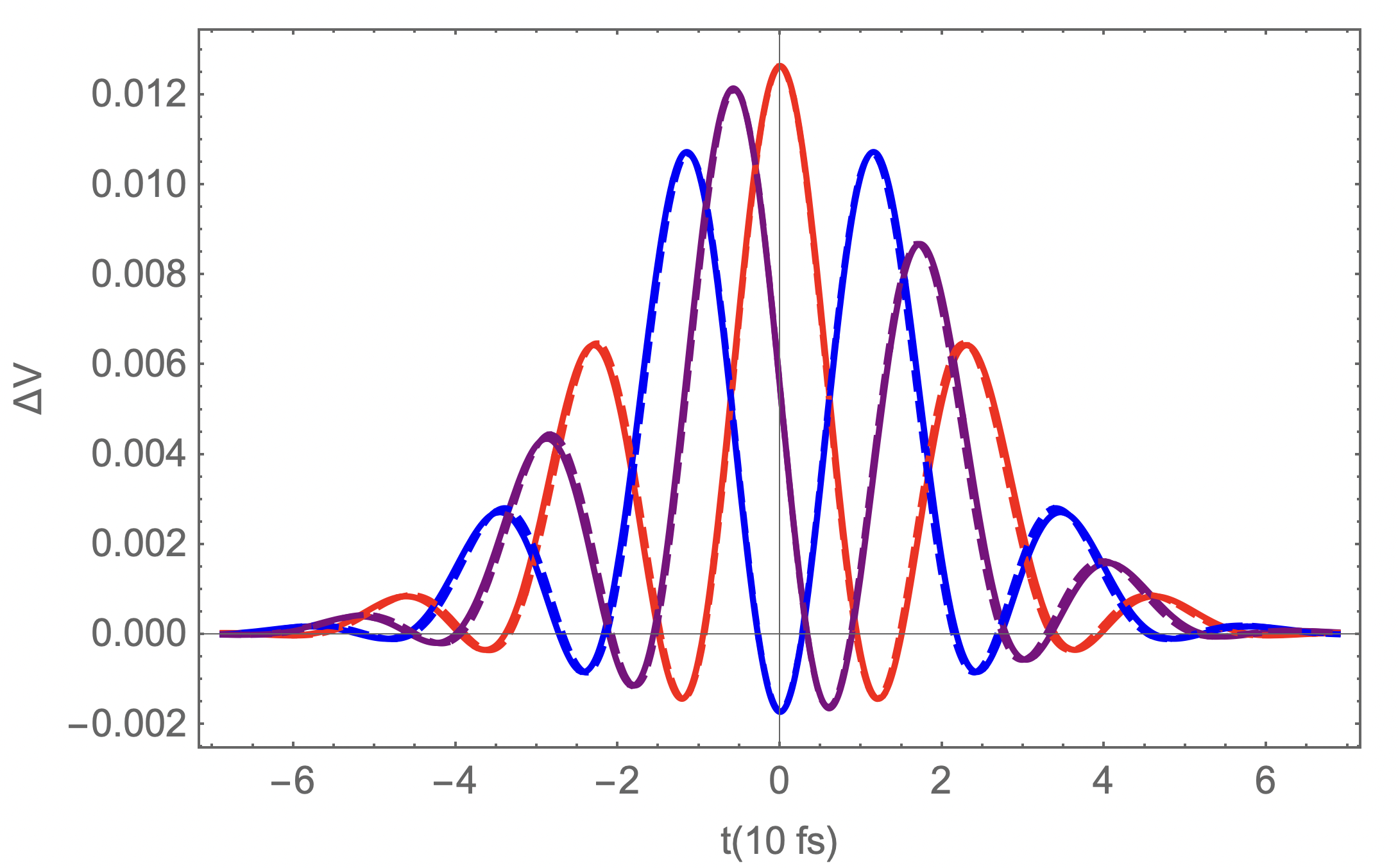}
\caption{The detected second moment (solid lines) are compared to the re-scaled second moment of the field (dashed lines). The red correspond to the electric-field quadrature, the blue correspond to the Hilbert quadrature and the purple correspond to their positive correlation.}
\label{FigSecondMoment}
\end{figure}

The sensitivity of the $n$th moment utilizing the standard EOS setup scales as $\epsilon^n$, where $\epsilon=\gamma_{E}$ is the coupling efficiency.
This makes it extremely challenging for EOS to be implemented for quantum sensing beyond the second moment.
However, a recent novel approaches to EOS utilizing a quantum probe has shown an enhancement to sensitivity to the higher moments \cite{StephaneQuantumEnhanced2021}.
For the purposes of extracting the information of a Gaussian state, it is sufficient to compute the second moment \cite{weedbrook2012gaussian, adesso2014continuous}, extracted through:
\begin{equation}
\begin{gathered}
\Delta V_{E}=(\braket{\hat{\textbf{E}}(t)^2}-\braket{\hat{\textbf{E}}(t)}^2)-(\braket{\hat{\textbf{E}}_0(t)^2}-\braket{\hat{\textbf{E}}_0(t)}^2)\,,
\\
\Delta V_{H}=(\braket{\hat{\textbf{H}}(t)^2}-\braket{\hat{\textbf{H}}(t)}^2)-(\braket{\hat{\textbf{H}}_0(t)^2}-\braket{\hat{\textbf{H}}_0(t)}^2) \, ,  
\end{gathered}
\end{equation}
where we have defined $\hat{\textbf{E}}_0$/ $\hat{\textbf{H}}_0$ as their respective operators $\hat{\textbf{E}}$/ $\hat{\textbf{H}}$ in the absence of the signal (i.e. in vacuum).
We utilize this set-up to analyze the statistics of a squeezed Gaussian signal:
\begin{equation}
\begin{gathered}
\hat{a}_{G}=\int_0^{\infty} \mathrm{d}\Omega \; G(\Omega) \hat{a}_{\Omega,Z}
\\
   G(\Omega)=B\sqrt{\omega}(\frac{1}{2 \pi \sigma_G^2})^{1/4}e^{(\Omega_0-\Omega)^2/(4 \sigma_G^2)}
\end{gathered}
\end{equation}
with $\Omega_0=20$ THz, $\sigma_G=4$ THz and a squeezing strength of $r=0.5$.
Our approach utilizing simultaneous measurement of both $\hat{E}_{BL}$ and $\hat{B}_{BL}$ allows the reconstruction of the Husimi Q-function (with added noise) in the time-domain, showing the correlation between the quadratures.
The detected second moment and their correlation are plotted in the Fig. \ref{FigSecondMoment}, and compared to the re-scaled second moment of the field.
The results demonstrate EOS's ability to conduct full tomography of a squeezed state at a relatively high accuracy.
The difference between the two results is associated with the imperfect mode-matching, which can be traced back to the use of the same probe-pulse for both the $\chi^{(2)}$ interaction and ellipsometry, and other technical effects such as phase-matching.

\section{Conclusion}
Our work establishes a method to conduct full tomography of a MIR squeezed state in the time-domain via EOS and introduces a quantitative method to analyse the quality of mode-matching, all of which are important milestones towards promoting EOS to a  time-domain quantum sensing technique.
Although the direct detection of $\hat{E}(t)$ and $\hat{H}(t)$ cannot be conducted in theory \cite{ScullyQuantumOptics, LoudonBook,MilburnQuantumOptics}, the quantum information of interest will exist within a certain bandwidth.
By introducing the same bandwidth to the electric field and its Hilbert transform quadrature, they are rendered detectable: $\hat{E}_{BL}(t)$ and $\hat{H}_{BL}(t)$.
Contrary to the standard approach to EOS which favors the use of a shorter probe pulse \cite{Matthias}, we have demonstrated that we can improve detection efficiency with the use of a longer probe-pulse, without the cost of accuracy.

The standard EOS set-up was shown to be very inefficient in coupling with the $\hat{H}_{BL}(t)$ field, and undetectable without an effective spectral filter \cite{Sulzer2020}.
We have created an efficient and accurate scheme to detect the Hilbert transform by introducing a phase-jump in frequency for the detection, and utilizing a delocalized probe field.
Furthermore, we achieved simultaneous detection of both $\hat{E}_{BL}(t)$ and $\hat{H}_{BL}(t)$ via multiplexing.
An experimental realization of this set-up would be the first example of full-tomography in the time-domain, and would be one of the most crucial advances in quantum sensing for MIR and ultra-fast photonics.

We found that a perfect phase-matching to $\hat{E}_{BL}(t)$ and $\hat{H}_{BL}(t)$ was not possible when the same probe pulse was utilized for driving the $\chi^{(2)}$ interaction and ellipsometry, especially in the presence of phase-matching.
For future research, implementing the use of template search \cite{Templatesearch} and a different probe-pulse for the $\chi^{(2)}$ interaction and ellipsometry as viable options to achieve perfect replication of the Husumi-Q function.

\twocolumngrid
\clearpage
\appendix
\section{Hilbert transform}\label{AppHilbertTransform}
In this section we present the effect of bandwidth on the coupling intensity, $\theta_H$, and mode-matching efficiency, $\gamma_H$, defined as:.

\begin{figure}
\includegraphics[width=\textwidth]{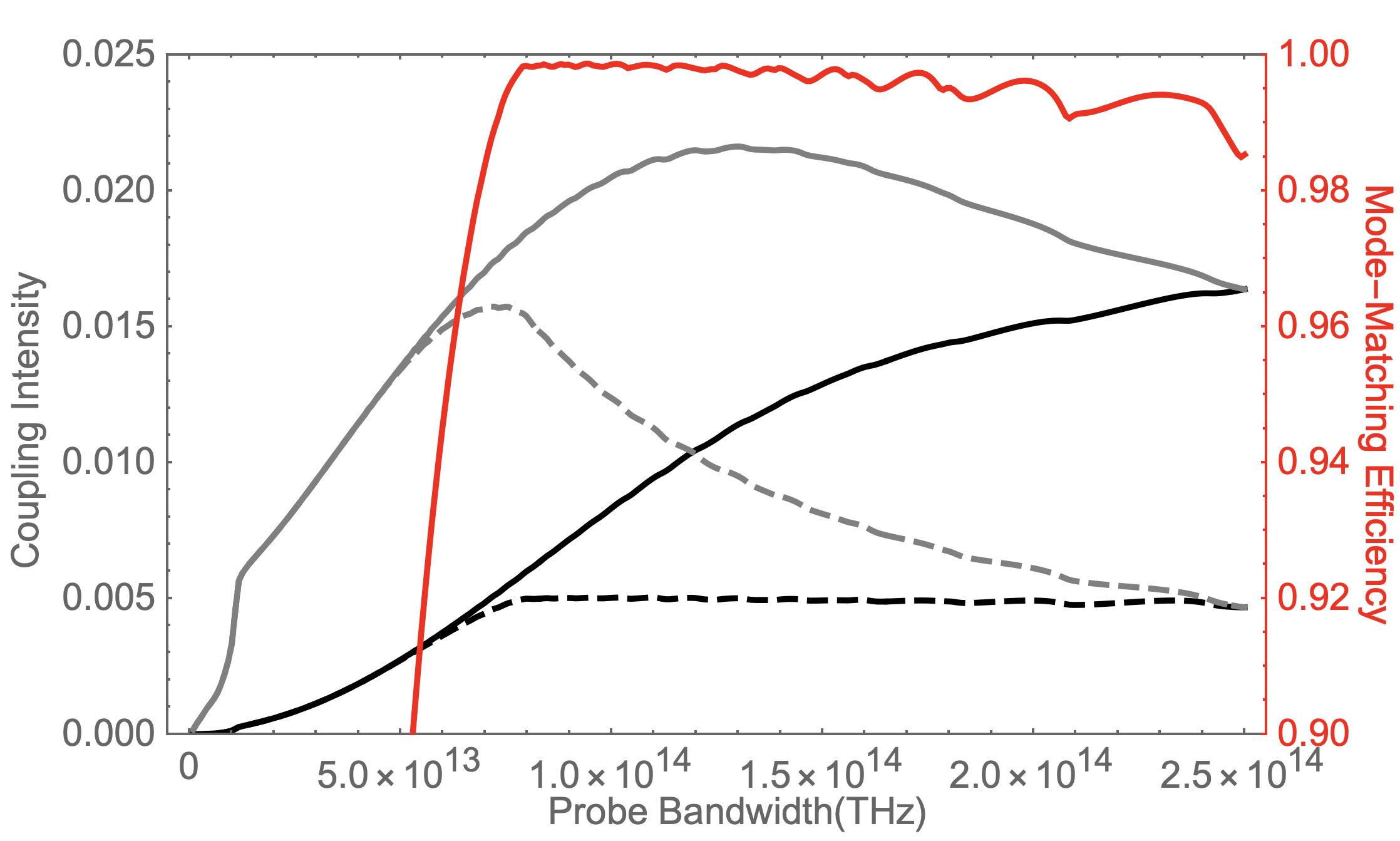}
\caption{We demonstrate the effect of bandwidth on coupling intensity and mode-matching efficiency. The black (dashed) line correspond to the constant photon number (bandwidth-limited) coupling intensity plots. The grey (dashed) line correspond to the constant intensity (bandwidth-limited) coupling intensity plots. The red line correspond to the bandwidth-limited mode-matching efficiency as a function of bandwidth.}
\label{FigDetectionEfficiencyHilbert}
\end{figure}

\begin{gather}
\theta_H=Tr_{\omega_m}(:[\hat{\mathbf{H}},\hat{\mathbf{H}}:)\, ,
\\
\gamma_{H}=Tr_{\omega_m}(:[\hat{a}_{H,S}',\hat{\mathbf{H}}]:)/\sqrt{\theta \theta_H}\, .
\end{gather}
Fig. \ref{FigDetectionEfficiencyHilbert} shows a similar trend to that of Fig. \ref{FigDetectionEfficiency}. For constant probe photon number, the bandwidth limited coupling intensity plateaus from a bandwidth of $80$ ThZ, giving a coupling intensity of $0.005$. The bandwidth-unlimited coupling intensity increases to $0.016$ when the full bandwidth is utilized, implying the extra bandwidth introduces unwanted noise. In comparison, for constant probe intensity, the bandwidth limited coupling intensity optimizes at around $80$ THz, with a coupling intensity of roughly $0.015$. This is a three fold increase in bandwidth-limited coupling intensity compared to when the full bandwidth was utilized. The mode-matching efficiency optimizes at about $99.9\%$ at around $80$ THz, then decreases for larger bandwidth due to phase-matching considerations.
\section{Experimental set-ups}\label{AppDifferentMethods}
The experimental method discussed in the main text allocated different bandwidths for the detection of the electric field and its Hilbert transform. Splitting the bandwidth does not come with any significant costs when considering a signal with $\Omega_0=20$THz, however for signals with large bandwidth, there is not enough bandwidth of the probe-pulse to allow multi-plexing. In this section, we discuss two alternate methods that can be implemented for detection of fields with larger bandwidths (signal with a central frequency of $25$ THz and above). 
\subsection{Beam-splitter method}\label{AppDifferentMethods1}
\begin{figure}[h]
\centering
\includegraphics[width=1\linewidth]{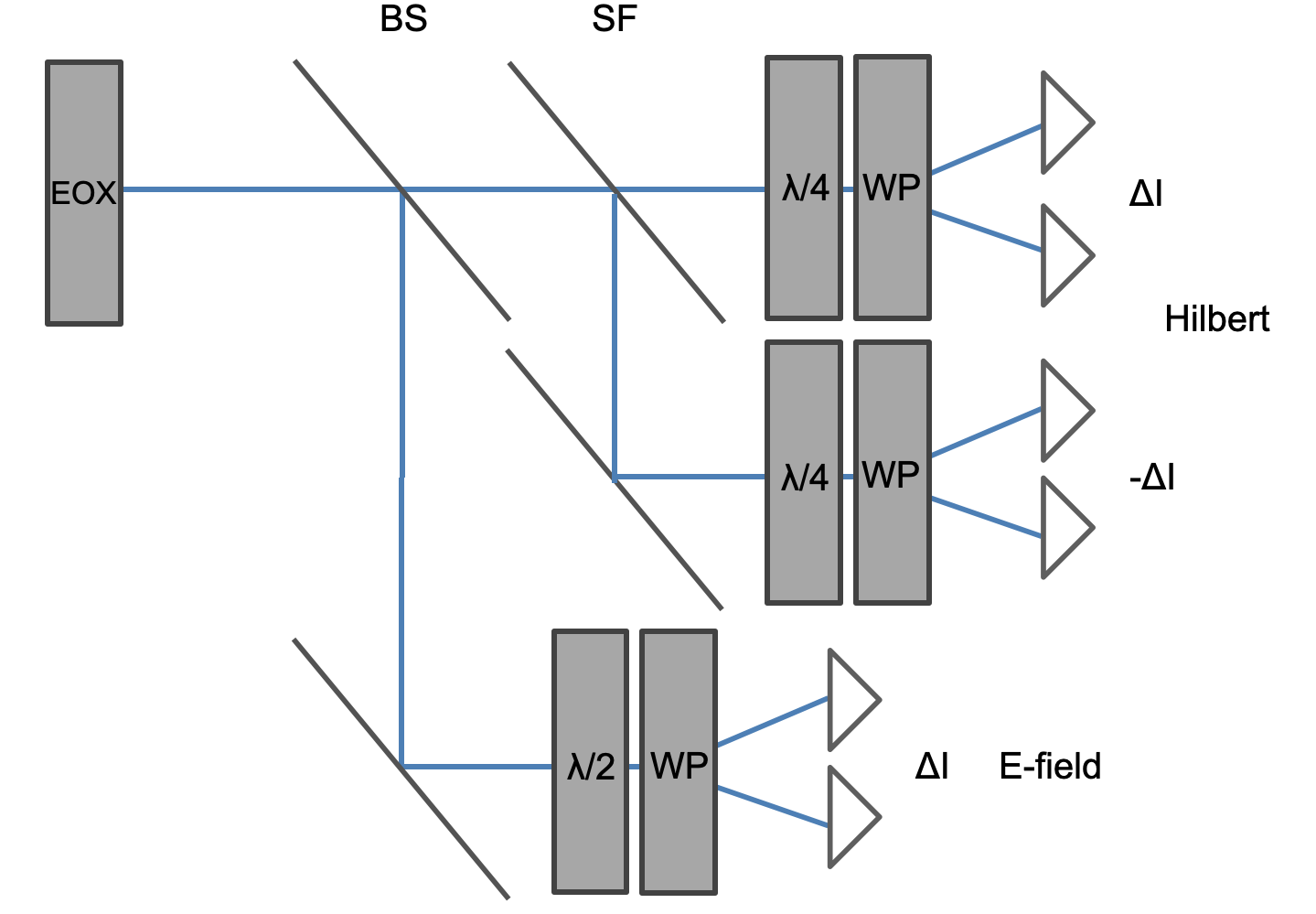}
\caption{Experimental set-up for simultaneous detection of both quadratures for signals with a larger bandwidth.}  \label{FigBSMethod}
\end{figure}

This method works by utilizing a probe pulse that is sufficiently efficient and accurate for both the detection of both the electric field and the Hilbert transform. In this paper, we suggest the use of a coherent superposition of the probe pulse to detect the electric field signal and the Hilbert transform signal. 
In general, the coupling to the Hilbert transform is stronger than the coupling to the electric-field. To account for this effect, we utilize a beam-splitter with a suitable transmission coefficient so that more photons are transmitted to the port extracting the Hilbert transform than the port extracting the electric field (e.g. if the coupling intensity for the electric field is $\theta_{E}=0.1$ and its Hilbert transform is $\theta_{H}=0.01$, then we utilize a beam-splitter which sends 10 times more photons in the Hilbert transform port). The electric field and Hilbert transform are simultaneously analysed in each port.  The experimental set-up is shown in Fig. \ref{FigBSMethod}.

\subsection{Tomography method} \label{AppDifferentMethods2}
There are several ways to conduct full tomography of the field. One is to extract the electric field and its Hilbert transform simultaneously. Another is to conduct an effective homodyne detection with various phases: an independent measurement of $\hat{X}(\phi)=\hat{a}e^{-i\phi}+\hat{a}^{\dag}e^{i\phi}$ with $\phi=\{0,\pi/2,\pi,3\pi/2\}$ is equivalent to a simultaneous measurement of $\hat{X}(\phi)$ with $\phi=\{0,\pi/2\}$. The prior can be achieved by utilizing a spectral filter to split the probe frequency above/below the central frequency into two ports. The outputs are followed by both quarter- and half-waveplate before the Wollaston prism. Careful tuning of both allows the detection of $\hat{X}(\phi)$ with an arbitrary phase. 

\begin{figure}[h]
\centering
\includegraphics[width=1\linewidth]{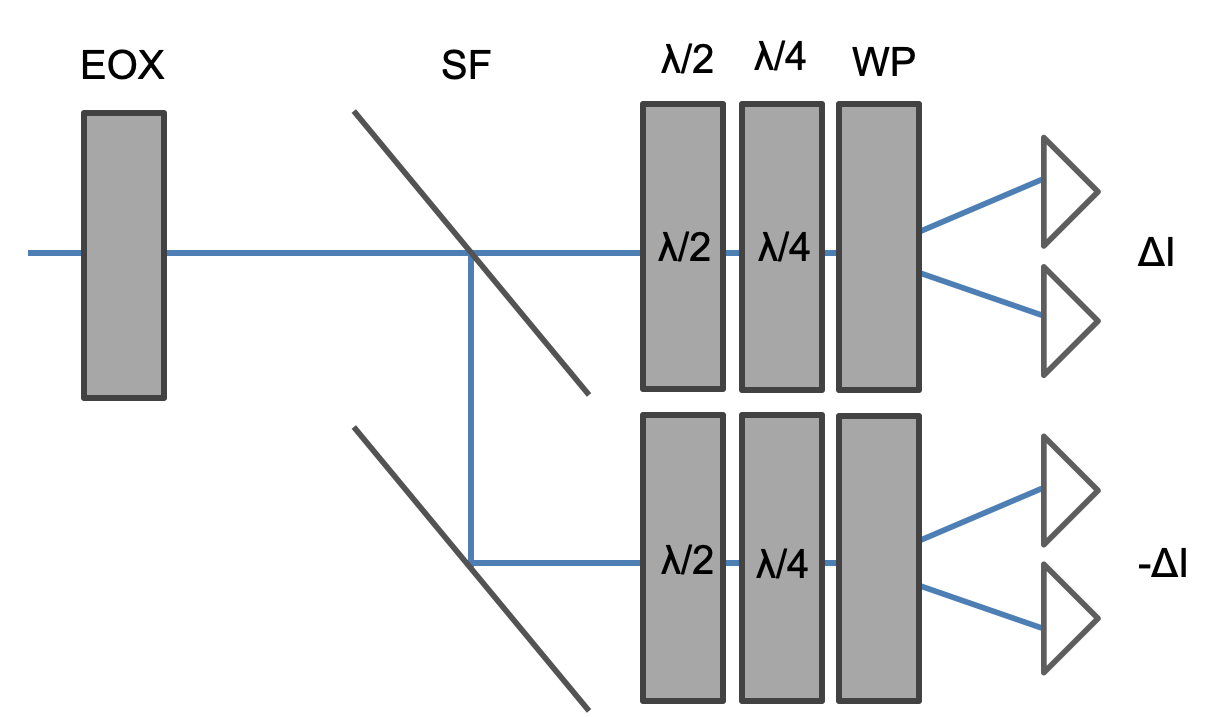}
\caption{Experimental set-up for sub-cycle weak homodyne detection, where the phase can be tuned via the angle of the half- and quarter-waveplates.}  \label{FigTMethod}
\end{figure}

\end{document}